\documentclass[12pt,a4paper]{article}

\usepackage[cp1250]{inputenc}
\usepackage[czech]{babel}

\def\uv#1{\raisebox{-1.3ex}[0ex][0ex]{``}\kern-.1ex#1\kern-.1ex''}

\begin{document}

\renewcommand{\thesubsection}{\arabic{subsection}}
\pagestyle{plain}
\setcounter{page}{1}

\begin{center}
\section*{GENERATING CONJECTURE AND SOME EINSTEIN-MAXWELL FIELD OF HIGH SYMMETRY}

M. {\v S}{\small{TEFAN{\'I}K}}, J. H{\small{ORSK{\'Y}}}

{\it{Department of Theoretical Physics and Astrophysics, Masaryk University,\\
Kotl{\'a}{\v r}ska 2, 611 37 Brno, Czech Republic}}
\end{center}

\vspace*{3ex}

{\small{\hspace*{3ex} For stationary cylindrically symmetric solutions of the Einstein-Maxwell
equation we have shown that the \uv{charged} solutions of McCrea, Chitre et al. (CGN), Van den
Bergh and Wils (VW) can be obtained from the seed metrics using generating conjecture. The McCrea
\uv{charged} solution has as a seed vacuum metric the Van Stockum solution with a Killing vector
$(0,0,1,0)$. The CGN \uv{charged} solution and the VW \uv{charged} solution have the static seed
metrics connected by the complex substitution $t \to iz$, $z \to it$ and the Killing vector which
is a simple linear combination of ${\partial}_{\varphi}$ and ${\partial}_{t}$ Killing vectors (VW),
respectively ${\partial}_{\varphi}$ and ${\partial}_{z}$  Killing vectors (CGN).}

\vspace*{3ex}

\subsection{Introduction}

\hspace*{3ex} There exist different methods to find the solutions of the Einstein--Maxwell equations
[1]. Some high symmetry solutions of the Einstein equations have direct astrophysical or
cosmological interpretation [2]. In [3], a new conjecture was formulated which says: \uv{A test
electromagnetic field having a potential proportional to the Killing vector of a seed vacuum
gravitational field (up to a constant factor), is simultaneously an exact electromagnetic potencial
of the self consistent Einstein--Maxwell problem where the metric is stationary or static and it
reduces to the mentioned seed metric when the parameter $K$ characterizing the charge of the
electromagnetic field source vanishes.}

\noindent For exact seed solution of the Einstein's equations we write

 \begin{equation}
  ds^{2}_{seed} = g_{ik}dx^{i}dx^{k}\;, \qquad i,k = 0,...,3.
  \label{1}
 \end{equation}

\noindent Let this metric has some symmetry expressed by the Killing vector field $\xi$. Acording to
the conjecture the electromagnetic four-potential $A_{i}$ is

 \begin{equation}
  A_{i}=K{\xi}_{i}=K\left ( g_{ik}{\xi}^{k}\right )\;.
  \label{2}
 \end{equation}

\noindent For the \uv{charged} solution of the Einstein--Maxwell equations we shall write

 \begin{equation}
  ds^{2}_{new}={\bar g}_{ik}dx^{i}dx^{k}\;.
  \label{3}
 \end{equation}

\noindent From (\ref{3}) we obtain the seed metric (\ref{1}) as a limit for $K \to 0$.

\noindent Let the \uv{charged} solution (\ref{3}) and the electromagnetic four-potential (\ref{2})
are known, we shall try to obtain the seed metric (\ref{1}) and the Killing vector field $\xi$.

\vspace*{3ex}

\subsection{Calculations}

\hspace*{3ex} Let us consider, first, the McCrea solution [4]. The nonzero \uv{charged} components
of it are

 \begin{equation}
  {\bar g}_{00}=4q^{2}r^{2}+C_{1}r{\ln \left ( kr\right )}\;,\qquad
  {\bar g}_{02}=-r\;,\qquad
  {\bar g}_{11}={\bar g}_{33}=-\frac{1}{\sqrt{r}}
  \label{4}
 \end{equation}

\noindent and the nonzero components of the seed metric ($q \to 0$ in (\ref{4})) are

 \begin{equation}
  g_{00}=C_{1}r{\ln \left ( kr\right )}\;,\qquad
  g_{02}={\bar g}_{02}\;,\qquad
  g_{11}=g_{33}={\bar g}_{11}={\bar g}_{33}\;.
  \label{5}
 \end{equation}

\noindent The electromagnetic four-potential of the McCrea's solution has the form

 \begin{equation}
  A_{i}dx^{i}=-qr\;dt\;.
  \label{6}
 \end{equation}

\noindent From (\ref{2}) we get

 \begin{equation}
  -qr=K\left ( g_{02}{\xi}^{2}\right )\;,\qquad
  g_{02}=-r\;.
  \label{7}
 \end{equation}

\noindent Taking $K = q$ we have the Killing vector

 \begin{equation}
  {\xi}^{i} \equiv \left ( 0, 0, 1, 0\right )
  \label{8}
 \end{equation}

\noindent which is just the Killing vector ${\partial}_{\varphi}$. The \uv{charged} solution [3] was
obtained by the conjecture for the Killing vector (\ref{8}) using the seed metric (\ref{5}). This
solution can be transformated to the McCrea's solution [5]. Therefore, we can formulate the
following theorem: \uv{The \uv{charged} McCrea's solution has as the seed metric the Van Stockum's
solution with the Killing vector field (\ref{8}).}

As the second example we will consider the \uv{charged} solution of CGN [4]. The nonzero components
of (\ref{3}) in the CGN's solution are

 \begin{equation}
  {\bar g}_{00}=r^{-4/9}{\exp{a^{2}r^{2/3}}}\;,
  {\bar g}_{02}=ar^{4/3}\;,
  {\bar g}_{22}=-\left ( r^{4/3}+a^{2}r^{2}\right )\;,
  {\bar g}_{11}=-r^{4/9}{\exp{a^{2}r^{2/3}}}\;,
  {\bar g}_{33}=-r^{2/3}.
  \label{9}
 \end{equation}

\noindent \uv{Charging} parameter of this solution is denoted in [4] as $a$, for the nonzero seed
metric coefficients we get

 \begin{equation}
  g_{00}=r^{-4/9}\;,\qquad
  g_{22}=-r^{4/3}\;,\qquad
  g_{11}=-r^{-4/9}\;,\qquad
  g_{33}=-r^{2/3}\;.
  \label{10}
 \end{equation}

\noindent The electromagnetic four-potential is

 \begin{equation}
  A_{i}dx^{i}=-\frac{a}{\sqrt{2}}r^{2/3}dz + \frac{a^{2}}{\sqrt{8}}r^{4/3}d{\varphi}\;.
  \label{11}
 \end{equation}

\noindent If we put $K = a$ in (\ref{2}) then the Killing vector of the seed metric (\ref{1}) is

 \begin{equation}
  \xi \equiv \left ( 0, 0, -\frac{a}{2}{\xi}^{3}, {\xi}^{3}\right )\;,\qquad
  {\xi}^{3}=\frac{1}{\sqrt{2}}\;.
  \label{12}
 \end{equation}

\noindent It is the linear combination of ${\partial}_{\varphi}$ and ${\partial}_{z}$ Killing
vectors.

Finally, let us consider the \uv{charged} solution of VW [4]. The nonzero metric coefficients of
(\ref{3}) are

 \begin{equation}
  {\bar g}_{00}=r^{2/3}\;,\quad
  {\bar g}_{02}=-ar^{4/3}\;,\quad
  {\bar g}_{22}=-\left ( r^{4/3}-a^{2}r^{2}\right )\;,\quad
  {\bar g}_{11}={\bar g}_{33}=-r^{-4/9}{\exp{-a^{2}r^{2/3}}}\;
  \label{13}
 \end{equation}

\noindent in the seed metric (\ref{1}) the following \uv{g's} are nonzero

 \begin{equation}
  g_{00}=r^{2/3}\;,\qquad
  g_{22}=-r^{4/3}\;,\qquad
  g_{11}=g_{33}=-r^{-4/9}\;.
  \label{14}
 \end{equation}

\noindent The line element (\ref{13}) has the electromagnetic four-potential in the form

 \begin{equation}
  A_{i}dx^{i}=-\frac{a}{\sqrt{2}}r^{2/3}dt+\frac{a^{2}}{\sqrt{8}}r^{4/3}d{\varphi}
  \label{15}
 \end{equation}

\noindent and it gives us the Killing vector

 \begin{equation}
  \xi \equiv \left ( {\xi}^{0}, 0, \frac{a}{2}{\xi}^{0}, 0\right )\;,\qquad
  {\xi}^{0}=-\frac{1}{\sqrt{2}}\;.
  \label{16}
 \end{equation}

\noindent This vector is the linear combination of ${\partial}_{\varphi}$ and ${\partial}_{t}$
Killing vectors.

The components of (\ref{14}) are coincident with (\ref{10}) by the complex substitution

 \begin{equation}
  t \to iz\;,\qquad z \to it\;.
  \label{17}
 \end{equation}

\noindent Both \uv{charged} solutions (\ref{13}) and (\ref{9}) have the similar electromagnetic
four-potential. Therefore, we can formulate the following theorem: \uv{The \uv{charged} metrics a)
CGN and b) VW can be obtained by the conjecture (\ref{2}) taking a) the static seed metric
(\ref{10}) and the Killing vector field (\ref{12}), respectively b) the static seed spacetime
(\ref{14}) and the Killing vector (\ref{16}).}

\vspace*{3ex}

\subsection{Conclusion}

\hspace*{3ex} The generating conjecture formulated in [3] can be used in both directions. If we
know the \uv{charged} static or stationary symmetric solution and the electromagnetic four-potential
then we can obtain the seed metric with the Killing vector field belonging to this seed metric. Two
theorem were established. The first one says: If we take the case II of the Van Stockum vacuum
solution [6] and the Killing vector (\ref{8}) then we obtain the \uv{charged} solution of McCrea. The
second theorem says: If we consider a) the static seed vacuum metric (\ref{10}) and the Killing
vector field (\ref{12}), respectively b) (\ref{14}) and (\ref{16}) then we can obtain a) the CGN
\uv{charged} solution, respectively b) the VW \uv{charged} solution.

The seed spacetimes (\ref{14}) and (\ref{10}) are connected by the complex substitution (\ref{17}).
Such considerations can be used in other cases as well.

\vspace*{3ex}

\begin{center}
{\large{\bf References}}
\end{center}

\noindent [1] Kramer D., Stephani H., MacCallum M.A.H., and Herlt E.: {\it Exact solutions of
                      Einstein's field equations.} VEB DAW, Berlin, 1980.\\
\noindent [2] Bi{\v c}{\'a}k J., Lynden-Bell D., and Pichon C.: Mon. Not. R. Astron. Soc. {\bf{265}} (1993) 126.\\
\noindent [3] Horsk{\'y} J., and Mitskievitch N.V.: Czech. J. Phys. B {\bf{39}} (1989) 957.\\
\noindent [4] MacCallum M.A.H.: J. Phys. A: Math. Gen. {\bf{16}} (1983) 3853.\\
\noindent [5] {\v S}tefan{\'\i}k M., and Horsk{\'y} J.: Scripta Masaryk University {\bf{24-26}} (1996) 39.\\
\noindent [6] Bonnor W. B.: J. Phys. A: Math. Gen. {\bf 13} (1980) 2121.

\end{document}